\documentstyle[preprint,aps]{revtex}

\newcommand{\be}{\begin{equation}}
\newcommand{\ee}{\end{equation}}
\newcommand{\n}{\nonumber}
\newcommand{\bea}{\begin{eqnarray}}
\newcommand{\eea}{\end{eqnarray}}

\begin{document}
\title{Coherent States for the Deformed Algebras}
\author{V. Sunilkumar, B. A. Bambah$^1$\thanks{e-mail:
bindu@panjabuniv.chd.nic.in},
P. K. Panigrahi\thanks{panisp@uohyd.ernet.in} and V.
Srinivasan\thanks{vssp@uohyd.ernet.in}}
\address{School of Physics,\\
University of Hyderabad,\\
Hyderabad, Andhra Pradesh,\\
500 046, India\\
     and}
\address{$^1$Department of Mathematics,\\
Centre for Advanced Study in Mathematics,\\
Panjab University,\\
Chandigarh,\\
160 014, India.}
\maketitle

\begin{abstract}
We provide a unified approach for finding the coherent states of various
deformed algebras, including quadratic, Higgs and q-deformed algebras,
which are relevant for many physical problems. For the non-compact
cases, coherent states, which are the eigenstates of the respective
annihilation operators, are constructed by finding the canonical
conjugates of these operators. We give a general procedure to map
these deformed algebras to appropriate Lie algebras. Generalized coherent
states, in the Perelomov sense,  follow from this construction. 
\end{abstract}
\newpage

    Recently, deformed Lie algebras have attracted considerable attention in
the context of various physical and mathematical problems\cite{tj}.
The quadratic algebra was discovered by Sklyanin\cite{skl1,skl}, in the context
of statistical physics and field theory. The well-known Higgs algebra, a
cubic algebra, was manifest in the study of the dynamical symmetries of the
quantum oscillator and the Coulomb problem in a space of constant
curvature\cite{le,higgs}. Other examples of deformation of Lie algebras
appear in the description of the degeneracy structures and as dynamical
symmetries of many conventional quantum mechanical problems,
like singular oscillators and Hartmann potential\cite{ze1,bon,vin,ze2}.
They have also appeared in interacting models of Calogero-Sutherland
type\cite{vin1,pani}. The presence of ambiguities in
the definition of the generators of the  Lie algebras, responsible for the
degeneracy in these problems, have led many authors to the deformed  Lie
algebras. The celebrated quantum groups is another example of deformation,
originating from the physical problems of the spin-chains and
lower dimensional integrable models\cite{drinj}.\\

In this communication, we present a unified approach for finding the
coherent states (CS) of these deformed algebras. 
Coherent states, needless to say,  occupy a very special place in physics, having
relevance to many problems of physical interest\cite{kla,perv}.
Hence, apart from its intrinsic interest, the method of construction
presented here, will greatly facilitate the physical applications of
these algebras. For the non-compact cases, the construction of the CS,
which are the eigenstates of the lowering operators, takes place in two
steps. First, we find the canonical conjugates of these operators.
The CS, corresponding to the deformed algebras are then obtained by
the action of the exponential of the respective conjugate operators on the
vacuum\cite{shanta,sc,}; this is in complete parallel to the harmonic oscillator
case. Another CS, which in a sense to be made precise in the text, is dual
to the first one, naturally follows from the above construction. We also
provide a mapping between the deformed algebras  and their undeformed
counterparts. This connection is then utilized to find the CS in the
Perelomov sense\cite{perv}. Apart from obtaining the known CS of the
$SU(1,1)$ algebra, we construct the CS for the quadratic, cubic and the
quantum group cases. Although our method is general, we will confine
ourselves here to finding the CS of the deformed $SU(1,1)$ and $SU(2)$
algebras.\\

The CS in our construction will be characterized by the eigenvalues of the
Casimir operator. This operator for the deformed algebra, written in the
Cartan-Weyl basis,
\be
[H\, , \,E_{\pm}]  = \pm E_{\pm}\,\,\,\,\,, \,\,\,\, [E_+\, , \,E_-]  =  f(H)\,\,\,\,,     
\ee
can be written in the form \cite{rocek}, 
\bea
C&=&E_- E_+ \,+\,g(H)\,\,\,, \n\\
 &=& E_+E_-\,+\,g(H-1)\,\,\,.
\eea
Here, $f(H)=g(H)-g(H-1),\,\,g(H)$ can be determined up to the addition of
a constant. The eigenstates are characterized by the values of the Casimir
operator and the Cartan
subalgebra $H$. We make use of these
relations to construct the canonical conjugate of the lowering operator of
the $SU(1,1)$ algebra and then write down the corresponding coherent
states. This is done for the purpose of comparing with the known results
in the literature and to illustrate our method. This
approach is then extended to the deformed algebras a straightforward way. 
In what follows, the relationships derived between various operators are
valid only on suitable Hilbert spaces.

For the $SU(1,1)$ algebra,

\be
   [K_+\, , \,K_-]  =  -2K_0  
\,\,,\,\,[K_{0} \,,\,K_{\pm}] =  \pm K_{\pm}\,\,\,\,, 
\ee
one finds, $f(K_0)= -2K_0$ and $g(K_0)=-K_0 (K_0 + 1)$. The quadratic
Casimir operator is given by $C=K_- K_+ + g(K_0) =K_-K_+ - K_0(K_0+1)$.
$\tilde{K_{+}}$, the canonical conjugate of $K_- $, satisfying
\be
[K_- \,,\,\tilde{K_+}]=1\,\,\,\,,
\ee
can be written in the form,
\be
\tilde{K_+}=K_+ F(C,K_0)\,\,\,\,.
\ee
Eq.(4) then yields,
\be
F(C,K_0)K_-K_+ - F(C,K_0-1)K_+K_- = 1\,\,\,\,;
\ee
making use of the Casimir operator relation given earlier, one can solve
for $F(C,K_0)$ in the form,
\be
F(C,K_0)=\frac{K_0 + \alpha}{C+K_0(K_0 +1)}\,\,\,\,.
\ee
The constant, arbitrary, parameter $\alpha$ in F can be determined by
demanding that Eq.$(4)$ is valid in the entire Hilbert space. For the
purpose of clarity, we illustrate this point, with the one oscillator
realization of the $SU(1,1)$ generators.

The ground states defined by $K_- \mid 0>=\frac{1}{2} a^2 \mid 0>=0$, are, 
  $\mid 0>$ and $\mid 1>=a^{\dag}\mid 0>$, in terms of the oscillator
Fock space. 
Making use of the results,
\be
K_0 \mid 0> =\frac{1}{4}(2a^{\dagger}a + 1) \mid 0> =\frac{1}{4} \mid 0>\,\,\,\,,
\ee
and
\be
C \mid 0>=\frac{3}{16}\mid 0>\,\,\,\,,
\ee
we find that, $[K_-\,,\,\tilde{K_+}]\mid 0>=K_-\tilde{K_+}\mid 0>$, yields $\alpha =\frac{3}{4}$. Similarly, for the other case
\be
[K_-\,,\,\tilde{K_+}] \mid 1>=\mid 1>\,\,\,\,,
\ee
leads to $\alpha = \frac{1}{4}$.
Hence, there are two disjoint sectors characterized by the $\alpha$ values
$\frac{3}{4}$ and $\frac{1}{4}$, respectively. These results match
identically with the earlier known ones \cite{shanta}, once we rewrite
$F$ as,
\be
F(C,K_0)= \frac{K_0 + \alpha}{C + K_0(K_0 + 1)}\,\,\,\,,
\ee
\be
 = \frac{K_0 + \alpha}{K_- K_+}\,\,\,\,.
\ee
The unnormalized coherent state $\mid\beta>$, which is the annihilation
operator eigenstate, i.e, $K_-\mid \beta >=\beta \mid \beta >$, is given
in the vacuum sector by
\be
\mid \beta> = e^{\beta \tilde{K^+}} \mid 0>\,\,\,\,.
\ee
Analogous construction holds in the other sector, where $\alpha=\frac{1}{4}$.
These states, provide a realization of the Cat states\cite{hill}, and play a
prominent role in the quantum measurement theory. As has
been noticed earlier\cite{shanta},  $[K_- \,,\,\tilde{K_+}]=1$, also yields,
\be
[\tilde{K_{+}^{\dagger}}\,,\,K_+]=1\,\,\,\,.
\ee
From this, one can find the eigenstate of $\tilde{K^{\dagger}_{+}}$
operator, in the form,
\be
\mid \gamma>=e^{\gamma K_+}\mid 0>\,\,\,\,.
\ee
This CS, after proper normalization is the well-known Yuen state\cite{yuen}.
Our construction can be easily generalized to various other realizations
of the $SU(1,1)$ algebra. \\

$\,\,\,\,$We now extend the above procedure to the quadratic algebra.
As has been mentioned earlier, this algebra has relevance to statistical
physics and field theory; a simpler version appears in quantum mechanical
problems:
\be
[N_0\,,\,N_{\pm}]=\pm N_{\pm}\,\,\,,\,\,\,[N_+ \,,\,N_-]=2N_0 + a N_{0}^{2}\,\,\,\,.
\ee
In this case, $f_1(N_0)=2N_0 + a N_{0}^2=g_1(N_0)-g_1(N_0 -1)$, where,
\be
g_1(N_0)=N_0(N_0 + 1)+\frac{a}{3}N_0(N_0+1)(N_0 + \frac{1}{2})\,\,\,\,. 
\ee
Representation theory of the quadratic algebra has been studied in the
literature\cite{rocek}; it shows a rich structure depending on the values
of `a'. In the non-compact case, i.e, for the values of `a' such that the
unitary irreducible representations (UIREP) are either bounded below or
above, we can construct the canonical conjugate $\tilde{N_{+}}$ of $N_-$
such that $[N_-\,,\,\tilde{N_+}]=1$. It is given
by $\tilde{N_+}=N_+F_{1}(C,N_0)$, with
\be
F_{1}(C,N_0)=\frac{N_0 + \delta}{C-N_0(N_0 +1)-\frac{a}{3} N_0(N_0+1)(N_0+\frac{1}{2})}\,\,\,\,.
\ee
As can be easily seen, in the case of the finite dimensional UIREP,
$\tilde{N_+}$ is not well defined since $F_1(C,N_0)$ diverges on the
highest state. As mentioned earlier, the values of $\delta$ can be fixed
by demanding that the relation, $[N_-\,,\,\tilde{N_+}]=1$, holds in the
vacuum sector $\mid v >_i$ where, $\mid v >_i$ are annihilated by $N_-$.
This gives $N_-\tilde{N_+}\mid v>_i=\mid v>_i$, which leads to
$(N_0+\delta)\mid v>_i=\mid v>_i$, the value of the Casimir operator,
$C=N_-N_+ + g_1(N_0)$, can be easily calculated.
Hence, the unnormalized coherent state $\mid \mu>_i: N_-\mid \mu>_i=
 \mu \mid \mu>_i$ is given by $e^{\mu \tilde{N_+}}\mid v>_i$.
The other coherent state originating from
$[\tilde{N_{+}^{\dagger}}\,,\,N_+]=1$ is given by $\mid \nu>_i=e^{\nu N_+}\mid v>_i$. This can be
recognized as the (unnormalized) CS in the Perelomov sense. Depending on the
UIREP being infinite or finite dimensional, this quadratic algebra can
also be mapped in to $SU(1,1)$ and $SU(2)$ algebras, respectively; leaving
aside the commutators not affected by this mapping, one gets,
\be
[N_+\,,\,\bar{N_-}] =- 2bN_0\,\,\,\,;
\ee
where $b=1$ corresponds to the $SU(1,1)$ and $b=-1$ gives the $SU(2)$ algebra.
Explicitly,
\be
\bar{N_-} = N_- G_1(C,N_0)\,\,\,\,,
\ee
and
\be
G_1(C,N_0) = \frac{(N_{0}^2 - N_0)b+\epsilon}{C-g_1(N_0 - 1)}\,\,\,\,,
\ee
$\epsilon$ being an arbitrary constant.
One can immediately construct CS in the Perelomov sense as $U\mid v>_i$, where
$U=e^{\xi N_+ -\xi^{\ast}N_-}$.
For the compact case, the CS are analogous to the spin and atomic coherent
states\cite{spin,atcs}.
We would like to point out that, earlier the generators of the deformed
algebra have been written in terms of the undeformed ones\cite{rocek}.
However, in our approach the undeformed $SU(1,1)$ and $SU(2)$ generators
are constructed from the deformed generators.

The Cubic algebra, which is also popularly known as the Higgs algebra in the
literature, manifested in the study of the degeneracy structure of
eigenvalue problems in a curved space\cite{higgs}.
The generators satisfy,
\be
[M_{0}\,,\,M_{\pm}] = \pm M_{\pm}\,\,\,,\,\,[M_{+}\,,\,M_{-}] = 2cM_0\,+\,4hM_{0}^3\,\,\,\,,
\ee
where, $f_2(M_0) = 2cM_0\,+\,4hM_{0}^3 = g_2(M_0)-g_2(M_0 - 1)$, and
\be
g_2(M_0) = cM_0(M_0 + 1)\,+\,hM_{0}^2(M_0 + 1)^2\,\,\,\,.
\ee
Analysis of its representation theory yields a variety of UIREP's, both finite
and infinite dimensional, depending on the values of the parameters
$c$ and $h$ \cite{zed}. Physically, $h$ represents the curvature of the
manifold. In the non-compact case the canonical conjugate is given by,
\be
\tilde{M_+} = M_+ F_{2}(C,M_0)\,\,\,\,,
\ee
where,
\be
F_{2}(C,M_0) = \frac{M_0 + \zeta}{C-cM_0(M_0+1)-hM_{0}^2(M_0 + 1)^2}\,\,\,\,.
\ee
As before, the annihilation operator eigenstate is given by
\be
\mid \rho>_i= e^{\rho\tilde{M_+}}\mid p>_i\,\,\,\,,
\ee
where, $\mid p>_i$ are the states annihilated by $M_-$. Like the previous
cases, the dual algebra yields another coherent state.  
This algebra can also be mapped in to $SU(1,1)$ and $SU(2)$ algebras, as
has been done for the quadratic case:
\be
[M_+\,,\,\bar{M_-}]=-2dM_0\,\,\,\,,
\ee
where, $d=1$ and $d=-1$ correspond to the $SU(1,1)$ and $SU(2)$ algebras respectively.
Here,
\be
\bar{M_-}=M_-G_2(C,M_0)\,\,\,\,,
\ee
where,
\be
G_2(C,M_0)=\frac{(M_{0}^2-M_0)d+\sigma}{C-g_2(M_0-1)}\,\,\,\,,
\ee
$\sigma$ being a constant.
The coherent state in the Perelomov sense is then $U\mid v>_i$, where,
$U=e^{\phi M_+ -\phi^{\ast}M_-}$.

For the sake of completeness, we now extend the above construction to the
quantum group case. The quantum deformed $SU(2)$ algebra is given by,
\be
[D_0\,,\,D_{\pm}] = \pm D_{\pm}\,\,,\,\,[D_+\,,\,D_-] = \frac{q^{D_0} - q^{-D_0}}{q-q^{-1}}\,\,\,\,, 
\ee
for which\cite{rocek},
\be
g_3(D_0) = \frac{q^{D_0 + \frac{1}{2}}\,+\,q^{-D_0 -\frac{1}{2}}}{(q^{\frac{1}{2}}
-q^{-\frac{1}{2}})(q-q^{-1})}\,\,\,\,.
\ee
The canonical conjugate $\tilde{D_+}$, of, $D_-$, valid for the non-compact
case, is 
\be
\tilde{D_+} = D_+F_{3}(C,D_0)\,\,\,\,,
\ee
where,
\be
F_{3}(C,D_0) = \frac{D_0 + \eta}{C-g_3(D_0)}\,\,\,\,.
\ee
One can then easily construct the coherent state like the previous examples. 
This algebra can also be mapped in to $SU(1,1)$ and $SU(2)$ algebras:
\be
[D_+\,,\,\bar{D_-}]=-2fD_0\,\,\,\,,
\ee
where, $f=1$ and $f=-1$ gives the $SU(1,1)$ and $SU(2)$ algebras
respectively (the other relations of the algebra not being affected in this
mapping).
Explicitly,
\be
\bar{D_-}=D_-G_3(C,D_0)\,\,\,\,,
\ee
where,
\be
G_3(C,D_0)=\frac{(D_{0}^2 -D_0)f+\omega}{C-g_3(D_0-1)}\,\,\,\,.
\ee
The coherent state in the Perelomov sense then follows naturally. 

To conclude, we have found a general method for constructing the coherent
states for various deformed algebras. Since the method is algebraic and
relies on the group structure
of Lie algebras, the precise nature of the non-classical behaviour
of these CS can be easily inferred from our construction. It will be of
particular interest to see the role of the deformation parameters in this
behaviour. Since many of these  algebras are related to
quantum mechanical problems with non-quadratic, non-linear Hamiltonians,
quantum optical problems involving 3-photon and four
photon processes  \cite{kara}, spin-chains and various other physical
problems\cite{deb,junk,vit}, a detailed study of
the properties of the CS associated
with these non-linear and deformed algebras is of physical relevance. 

The authors take the pleasure to thank Prof. S. Chaturvedi for stimulating
conversations. VSK acknowledges useful discussions with Mr. N. Gurappa.

\end{document}